\documentclass{Earthzine}

\newcommand{\EZyear}{2014}
\newcommand{\EZvolume}{7}

\newcommand{\EZissue}{2}
\newcommand{\EZissueText}{$2^{nd}$ quarter theme. Geospatial Semantic Array Programming}

\EZcomment{}
\newcommand{\EZpage}{910137}

\expandafter\def\expandafter\UrlBreaks\expandafter{\UrlBreaks
  \do\a\do\b\do\c\do\d\do\e\do\f\do\g\do\h\do\i\do\j%
  \do\k\do\l\do\m\do\n\do\o\do\p\do\q\do\r\do\s\do\t%
  \do\u\do\v\do\w\do\x\do\y\do\z\do\A\do\B\do\C\do\D%
  \do\E\do\F\do\G\do\H\do\I\do\J\do\K\do\L\do\M\do\N%
  \do\O\do\P\do\Q\do\R\do\S\do\T\do\U\do\V\do\W\do\X%
  \do\Y\do\Z}

\title{Estimating the effects of water-induced shallow landslides on soil erosion}

\newcommand{\EZsearch}{Predicting+Future+Forest+Ranges+Using+Array-based+Geospatial+Semantic+Modelling}
\EZregisterTitle

\EZauthor{1}{Claudio}{C.}{Bosco}
\EZauthor{1}{Graham}{G.}{Sander}
\EZaffiliation{1}{Loughborough University, Department of Civil and Building Engineering Loughborough,~United~Kingdom\vspace{2mm}}

\EZregisterAuthors

\EZregisterAbstract{
Rainfall induced landslides and soil erosion are part of a complex system of multiple interacting processes, and both are capable of significantly affecting sediment budgets. These sediment mass movements also have the potential to significantly impact on a broad network of ecosystems health, functionality and the services they provide. To support the integrated assessment of these processes it is necessary to develop reliable modelling architectures. This paper proposes a semi-quantitative integrated methodology for a robust assessment of soil erosion rates in data poor regions affected by landslide activity. It combines heuristic, empirical and probabilistic approaches. This proposed methodology is based on the geospatial semantic array programming paradigm and has been implemented on a catchment scale methodology using Geographic Information Systems (GIS) spatial analysis tools and GNU Octave. The integrated data-transformation model relies on a modular architecture, where the information flow among modules is constrained by semantic checks. In order to improve computational reproducibility, the geospatial data transformations implemented in Esri ArcGis are made available in the free software GRASS GIS. The proposed modelling architecture is flexible enough for future transdisciplinary scenario analysis to be more easily designed. In particular, the architecture might contribute as a novel component to simplify future integrated analyses of the potential impact of wildfires or vegetation types and distributions, on sediment transport from water induced landslides and erosion.
}

\usepackage{ulem}

\newcommand{\checkisnote}[1]{\href{http://mastrave.org/doc/mtv\_m/check\_is\#SAP\_#1}{\bf{::#1::}\footnote{\scriptsize \url{http://mastrave.org/doc/mtv\_m/check\_is\#SAP\_#1}}}}

\begin{document}
\EZmaketitle

\section{Introduction}

Hillslope processes can be envisaged as a cascade where surface erosion and mass movements are visible expressions of critical instabilities in a complex system of interacting processes that control the downslope movement of material \cite{Van_Asch_1980} in \cite{Van_Beek_2002}. Field observations, modelling simulations and experimental studies have shown that soil erosion can vary considerably due to the changes in soil properties, vegetation cover and topography occurring after a landslide (e.g. \cite{Cochrane_Acharya_2011,Acharya_etal_2009,Acharya_etal_2011}). Following landslide events the changes in soil erosion rates can be strong enough to deliver significant cascading impacts on ecosystems, for example due to an increased sediment yield to a stream network. This may potentially be of ecological and economical relevance not only locally (possibly driving complex changes even at the landscape-scale \cite{Bakker_etal_2005,Geertsema_Pojar_2005}) but also off-site, whenever ecosystem services are important for service benefit areas connected through service connecting areas \cite{Syrbe_Walz_2012} (e.g. stream networks).

As natural resources are intrinsically entangled in complex networks there is a growing awareness of the importance of these cascades.  This, in turn is driving the development of integrated risk assessment and multi-purpose use optimization of different resources to develop appropriate management policies that can reliably model the potential influence of climate change on these process cascades, and assess the resultant economic and societal consequences.

Landslide events will result in changes in topography and vegetation cover which in turn will alter surface erosion rates and sediment yields.   There are a number of relevant models that use an integrated approach to soil erosion and landslide processes, including SHETRAN (the name derived from Système Hydrologique Européen-TRANsport) \cite{Ewen_etal_2000}, TOPOG (a physically-based, distributed parameter, catchment hydrological model) \cite{O_Loughlin_1986,CSIRO_2012}, PSIAC (Pacific Southwest Inter-Agency Committee) \cite{PSIAC_1968} or SIBERIA (also known as the Willgoose Catchment Evolution Model) \cite{Willgoose_Riley_1998}. WEPP-SLIP (Water Erosion Prediction project - Shallow Landslide Integrated Prediction) \cite{Cochrane_Acharya_2011} is a model that explicitly considers post-failure sediment yield. This model integrates the physical basis of the WEPP model \cite{Laflen_etal_1991}, with the infinite slope stability model of Skempton and DeLory \cite{Skempton_DeLory_1957}. WEPP-SLIP is able to consider the post-failure changes in soil erosion rate through the changes in topography and land cover.

Physically based models use a dynamic hydrological approach and local terrain characteristics for estimating spatial and temporal landslide probability \cite{Jaiswal_Van_Westen_2009}. The main limits of physically based models are that they are often optimised for small catchments and local conditions, and that these require in depth knowledge of local soil and climatological parameters \cite{de_Vente_etal_2013}.  Empirical methods are mainly based on the estimation of thresholds related to precipitation patterns which result in landslide occurrence \cite{Jaiswal_Van_Westen_2009}. This approach generally requires high temporal resolution rainfall data, which is not often available, and does not necessarily model the right processes. In addition it is limited to being applicable to only the same conditions under which it was developed \cite{Hessel_2002,de_Vente_etal_2013}.  However, there is still room to improve the modelling of the interactions of these processes, for example through assessments of the changes in surface area made more susceptible to soil erosion following landslide events.

To quantify the potential changes in soil erosion due to landslide occurrence it is necessary to know where and when on the slope a landslide initiates and how it evolves. This paper aims to present a new modelling approach for data-poor regions in an attempt to improve the estimation of sediment budgets derived from rainfall induced landsliding and soil erosion. A statistical approach is proposed that is based on incorporating the frequency-area landslide distribution model of Malamud et al. \cite{Malamud_etal_2004} within the framework of a spatially distributed empirical soil erosion model.

\section{the study area}

The study area (Fig.1) is situated in southern Italy in the Daunia Appennines of the Puglia region, within the municipal territory of Rocchetta Sant’Antonio. It covers an area of almost 10 $km^2$. This area is highly susceptible to landslide activity \cite{Iovine_etal_1996,Magliulo_etal_2008} with a consequent negative impact on the local economy \cite{Wasowski_etal_2010}. The neighbouring area to the north-west of the Rocchetta Sant'Antonio territory presents a landslide frequency exceeding 20\% for the overall area \cite{Mossa_etal_2005,Wasowski_etal_2007,Wasowski_etal_2010,Wasowski_etal_2012}. Soil erosion is also widespread and the severity is largely determined by the combination of tillage practices and the high erodibility of the clay-rich flysch units from which some of the local soils are derived \cite{Lamanna_etal_2009}.  Within the catchment it is possible to distinguish four major classes of land use (agricultural soils, woodland, pastures and grassland) and three dominant lithologies (limestone, sandstone and clay). Slope angles are on average approximately 10 degrees with peak slope angles rarely exceeding 25 to 30 degrees. An ephemeral drainage network is fed by precipitation during the autumn-winter period when some 600 to 750 mm of rainfall is common \cite{Wasowski_etal_2010}. The area is characterized by a Mediterranean sub-humid climate.

\begin{figure}
\vspace{-1mm} 
\centerline{\includegraphics[scale=0.5]{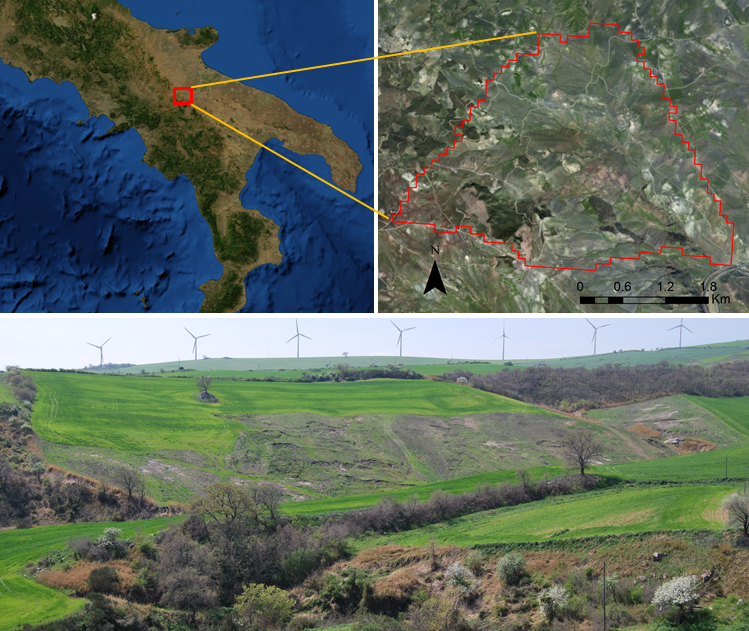}}
\caption{Figure 1: The study area (Rocchetta Sant'Antonio, Italy). Google Earth, \copyright 2013 Google.}
\vspace{-3mm} 
\end{figure}

\section{A new architecture for coupling of the effects of rainfall-induced shallow landslides and soil erosion}

\subsection{geospatial semantic array programming}

Array programming is an approach for simplifying complex algorithm prototyping with an accurate and compact mathematical description. It originates as a means for reducing the gap between mathematical notation and its implementation within the model’s algorithms in a formalised and reproducible way. As stated by Iverson \cite{Iverson_1980}: “the advantages of executability and universality found in programming languages can be effectively combined, in a single coherent language, with the advantages offered by mathematical notation”.  Array programming has been used for building the architecture for our modelling approach. For mitigating the complexity of trans-disciplinary modelling and the inconsistencies between input data, parameters and output, semantic checks on the processed information and a modularisation of the key parts of the model were introduced following the semantic array programming paradigm (SemAP) \cite{de_Rigo_etal_2013,de_Rigo_2012a,de_Rigo_2012b}. The proposed architecture (Fig. 2) exploits the geospatial capacities of GIS in order to estimate soil erosion yield (e-RUSLE model).  In our approach we integrated SemAP and geospatial tools (ArcGis and GRASS GIS) through the Geospatial Semantic Array Programming paradigm (GeoSemAP). GeoSemAP exploits geospatial tools and Semantic Array Programming for splitting a complex data-transformation-model (D-TM) into logical blocks whose reliability can more easily be checked by applying geospatial and mathematical constraints.

Semantic checks are exemplified in the following paragraphs with the notation \textbf{::constraint::}. The semantic constraints were implemented within the code with a specialised module \cite{de_Rigo_2012c} of the Mastrave modelling library. A hyperlink to the corresponding online description is provided.

\begin{figure}
\vspace{-1mm} 
\vspace{1 cm} 
\centerline{\includegraphics[scale=1.10]{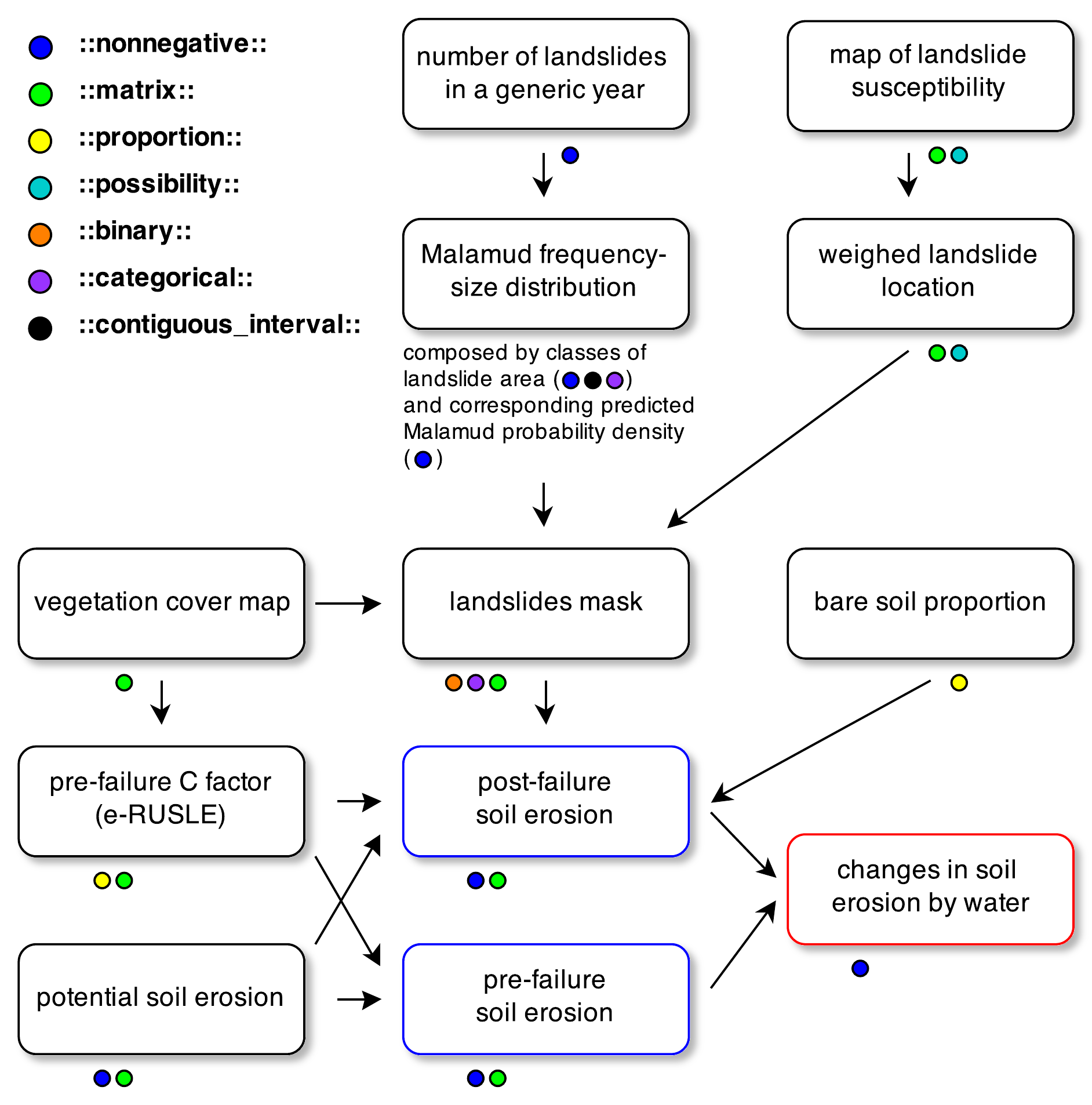}}
\caption{Figure 2: Flowchart of the model. The semantic aspects of the data-transformations among model components are highlighted within the workflow.}
\vspace{-3mm} 
\end{figure}

\subsection{applied techniques}

The pre- and post-failure soil loss rate was calculated by applying the low data demanding model e-RUSLE \cite{Bosco_etal_2014}. This model retains all the equations of its predecessor (RUSLE, \cite{Renard_etal_1997}) and implements an extra factor to account for the effects of soil stoniness on soil erosion. Due to the flexibility of the modelling architecture that e-RUSLE is based on, it is possible to calibrate the model for application at different scales \cite{Bosco_etal_2014}. e-RUSLE was implemented using the ArcGIS software to first estimate the \checkisnote{nonnegative} \checkisnote{matrix} representing the soil erosion rates within  the catchment without considering the influence of mass movement. The scripts applied for calculating the soil erosion losses can also be easily carried out using an Open Source Free Software such as GRASS GIS or Quantum GIS.

To determine the slope length factor required in e-RUSLE, the D-infinity (D$\infty$) algorithm of Tarboton \cite{Tarboton_1997} was first used to calculate the flow direction and then the flow length.  Due to the geomorphological characteristics of the study area, a multiple-neighbour flow algorithm was required with the D$\infty$ algorithm being one of the most suitable \cite{Gruber_Peckham_2009,Chirico_etal_2005,Erskine_etal_2006}. In GRASS GIS it is possible to apply a multiple-flow approach using the tool 'r.watershed' \cite{Ehlschlaeger_2014}. The slope steepness factor was also slightly modified in comparison to the application of the e-RUSLE presented in Bosco et al. \cite{Bosco_etal_2014}. This was based on the Nearing’s \cite{Nearing_1997} equation which performs best for higher slopes \cite{Bosco_etal_2008,Bosco_etal_2014}.  However the Moore and Burch \cite{Moore_Burch_1986} formula is more appropriate for slopes lower than 12.73 degrees because it gives the correct limiting value of zero in absence of any steepness.  A comparison of both formulas is presented in Fig. 3, where a close matching trend is observed between 0 and 12.73 degrees (or 0 - 0.22 rad).  Consequently a merged formula can be obtained by using the Moore and Burch equation for slopes less than 12.73 degrees and then the Nearing formula for higher slopes. To calculate the slope steepness factor of the model, the tool r.slope.aspect \cite{Shapiro_Waupotitsch_2014} of GRASS can be used. The majority of the equations that e-RUSLE is based up have been applied using the ArcGis tool ’Map Algebra’ that in GRASS corresponds to ’r.mapcalc’ \cite{Shapiro_Clements_2014}.

\begin{figure}
\vspace{-1mm} 
\centerline{\includegraphics[scale=0.55]{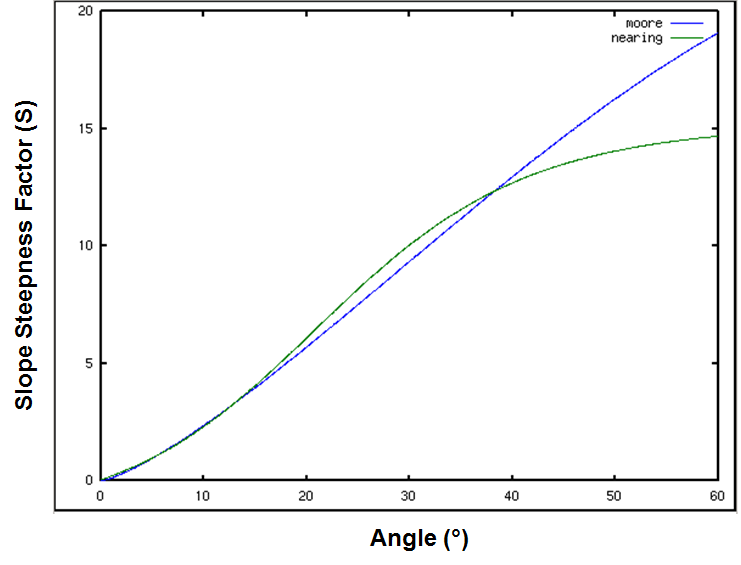}}
\caption{Figure 3: Comparison between the Moore and Burch \cite{Moore_Burch_1986} relation and the Nearing’s \cite{Nearing_1997} formula applied for calculating the S factor of the e-RUSLE model.}
\vspace{-3mm} 
\end{figure}

For quantifying the effect of size, position and number of landslides affecting this catchment the frequency-size distribution model proposed by Malamud et al. \cite{Malamud_etal_2004} was adopted.  They found that landslide data from three quite different locations around the world (Italy, Guatemala and the United States) could be described quite well with the inverse gamma distribution  

\begin{equation}
\label{inverse-gamma distribution}
p(A_{L}, \rho, a, s) = \frac{1}{a\Gamma(\rho)}\left [ \frac{a}{A_L-s} \right ]^{\rho+1} \exp \left [ -\frac{a}{A_L-s}\right ]
\end{equation}
\vspace{3mm}

In (1), $p$ = probability density ($km^{-2}$), $\Gamma$ is the gamma function, $A_L$ = the landslide area ($km^2$), $\rho$ (-) is a parameter which controls the power law decay for medium and large landslide areas, $a$ ($km^2$) determines the position of the maximum in the probability distribution  and $s$ ($km^2$) is a parameter which fits the exponential decay behaviour for small landslide areas.	Parameter values of $\rho$ = 1.4, $a$ = 1.28 $10^{-3}\; km^2$ and $s$ = -1.32 $10^{-4}\; km^2$ were shown to provide a good fit to the measured data.									
A dataset of more than 400 reported landslides that affected the catchment in 2006 was made available and published by Dr Janusz Wasowski of CNR-IRPI, Bari \cite{Wasowski_etal_2010,Wasowski_etal_2012}. For obtaining the landslide inventory, high resolution IKONOS satellite imagery was used. To make the interpretation easier, the satellite images were orthorectified and pansharpened.
This dataset is not freely available but the IFFI (Inventario dei Fenomeni Franosi in Italia) database \cite{Agnesi_etal_2007} is a valuable alternative to apply our modelling approach whenever enough data are available.

Overall a reasonable correlation between the inverse-gamma distribution of Malamud et al. \cite{Malamud_etal_2004} with the above parameter values and the frequency-size distribution of the landslide database was found (Fig. 4).  The fit is very good for landslide areas greater than or equal to the peak in the distribution.  For smaller landslide areas to the left of the peak the agreement is not as good, though modifications to parameters a and s could be made to improve this section. However the distribution of Malamud et al. \cite{Malamud_etal_2004} and parameter values they used, were shown to work over a wide range of landslide sizes from various countries around the world.  It was found that these same parameter values also provided a similar fit to the data from our field site suggesting the possibility of universality in the parameter values and therefore removing the need for calibrating the distribution for local applications.  On this basis we wanted to see how well this would perform against data from the Rocchetta catchment and kept the original Malamud parameter values. The data for the smaller landslides does have a greater degree of uncertainty as its collection could easily have led to either an over or underestimation of the landslide number.  This could occur through either medium landslides being classified as smaller due to being covered by larger landslides, or though the smaller landslides being covered by larger ones and therefore missed completely.
The main point of this exercise wasn't to match exactly the landslide-area probability distribution, but to have a physically realistic distribution on which to base our modelling. 
To predict when and where a landslide will occur is one of the main challenges for calculating post-failure soil loss in data-poor regions. We exploited the correlation between the measured data and Malamud’s distribution through combination with Monte Carlo simulation to analyse the effects of mass movements on soil erosion by water.

\begin{figure}
\vspace{-1mm} 
\centerline{\includegraphics[scale=0.55]{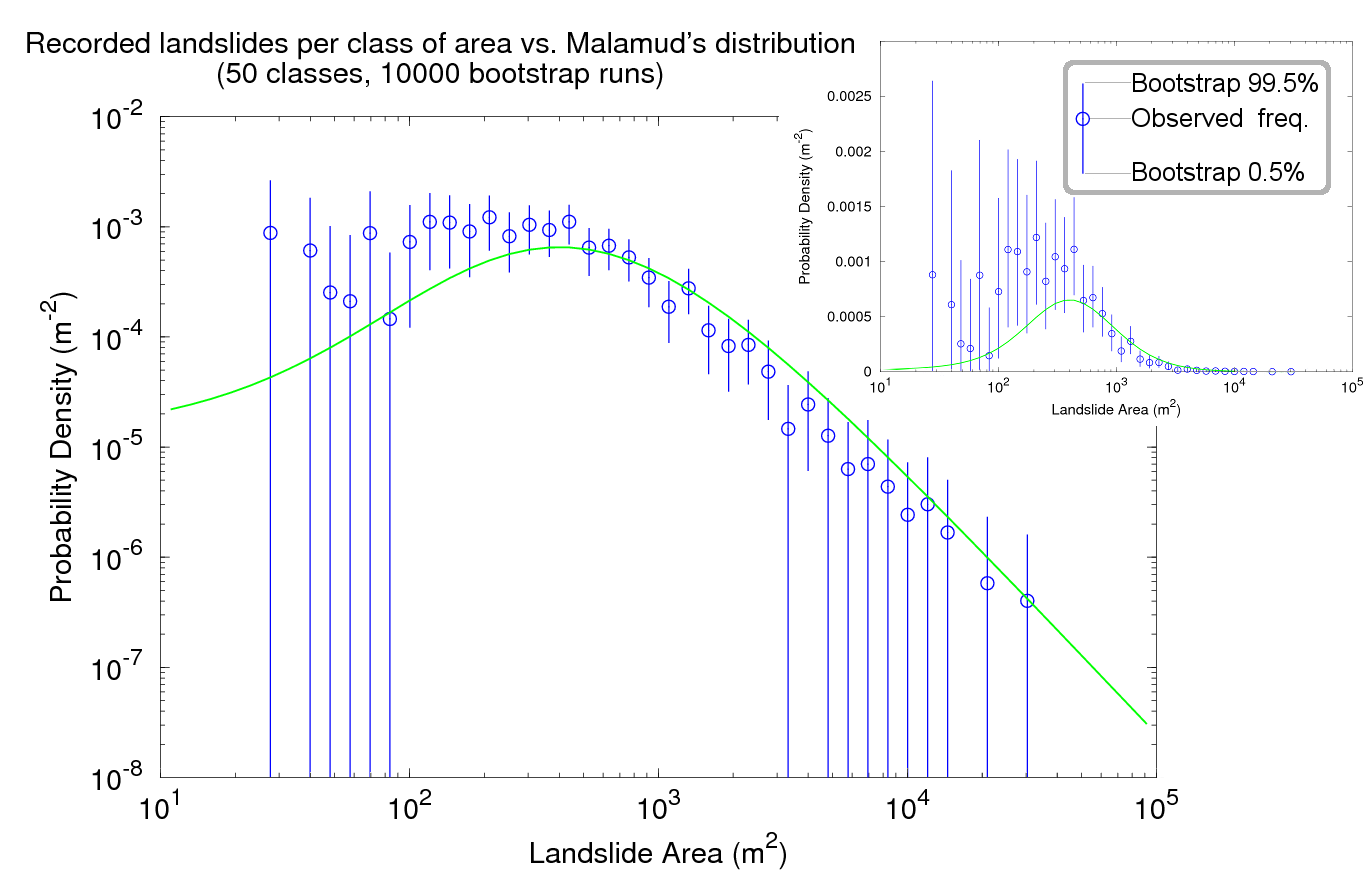}}
\caption{Figure 4: Dependence of the landslide probability densities on landslide area for the measured set of data (blue) and for Malamud’s distribution (green). The probability density is given on logarithmic and semi-logarithmic scale. A bootstrap analysis was performed to assess the uncertainty of the measured data.}
\vspace{-3mm} 
\end{figure}

Assuming the validity of the proposed inverse-gamma function for calculating the probability distribution of landslide areas we implemented a simple script (based on SemAP) in MATLAB language. Starting from a \checkisnote{scalar\_positive} number to represent the number of landslides that occurred in the catchment, we then calculate the number of landslides $\delta N_L(h)$ in the h-th class of landslides. Each class is a \checkisnote{categorical-interval} which includes all the landslides with an area from $A_L (h)$ to $A_L (h + 1)$. The classes thus form a partition of \checkisnote{contiguous\_interval} $s$ in $[0,A_L (hmax )]$  whose values are found from:

\begin{equation}
\label{integral}
\delta N_L(h) = \int_{A_L(h)}^{A_L(h+1)} p(A_L)\, dA_L
\end{equation}
\vspace{3mm}

In order to evaluate the effect of the post-failure changes on the soil erosion rates in the catchment, we applied the Monte Carlo method twice. Once to randomly determine the location of a landslide and a second time to sample the Malamud distribution to assign its size. The Monte Carlo simulation was also implemented in the MATLAB language following the SemAP paradigm and exploiting the potentiality offered by the Mastrave Library \cite{de_Rigo_2012a} whose tools were largely used within the code. 

To be more explicit: considering $Y$ as a random variable distributed according to a given probability distribution, it is possible to generate n pseudo-random instances $Y_1$,..., $Y_n$ with the same distribution . This may be accomplished with a classical Monte Carlo extraction. Let us define $f(\cdot)$ as a certain function of $Y$ which is implemented, within the SemAP paradigm, as a D-TM transforming an instance of $Y$ into the desired output data. Suppose we are interested in computing the integral A of $f(\cdot)$ over a given domain $\Omega$. This implies considering the probability density function $\pi(\cdot)$ of $Y$ over $\Omega$:

\begin{equation}
\label{eq:theoreticalIntegral}
A = \int_{\Omega} f( Y ) \cdot \pi(Y)\,dY, \qquad {\begin{array}{l}
Y \in  \Omega \\
Y \sim \Phi \\
\pi(Y) \text{ density function of $\Phi$ in $Y$} \\
\text{such that } \displaystyle \int_{\Omega} \pi(Y)\,dY = 1
\end{array}}
\end{equation}
\vspace{3mm}

\newpage
Numerically, it is possible to approximately estimate $A$ by exploiting the $n$ Monte Carlo instances $Y_1$,..., $Y_n$ as

\begin{equation}
\label{inverse-gamma distribution}
A \approx \hat{A_n} = \frac{1}{n} \sum_{\text{run}=1}^n f(Y_\text{run}), \qquad \forall \,\text{run}, Y_\text{run} \sim \Phi
\end{equation}
\vspace{3mm}

where $Y_{run}$ is the $run$-th instance of $Y$ corresponding to the $run$-th Monte Carlo iteration. From the law of large numbers, if $n \Rightarrow \infty, \hat{A}_n \Rightarrow A$. In our particular application, $\hat{A}_n$ is the average over n runs of simulated landslides; in each of them the total erosion by water $f(\cdot)$ is computed for the particular array of landslides $Y_{run}$ . The n arrays of simulated landslides are the basis for $f(\cdot)$ to estimate the corresponding post-landslide soil erosion. Each landslide occurring in the $run$-th simulation has an area distributed according to $\bar{p}(\cdot)$. This defines $\pi(\cdot)$ as the probability density function with  which each $run$-th array of landslides is distributed.

The Monte Carlo simulation was iterated 1,000 times. For each of the iterations the post-failure changes in soil erosion were calculated and compared with the pre-failure estimates.

The \checkisnote{matrix} representing the cover management factor of the e-RUSLE model was calculated using a 5x5 metres resolution land cover map of the study site, produced by CNR-IRPI of Bari using ASTER satellite multi-spectral imagery and published in \cite{Wasowski_etal_2010}. The map is not freely available but the CLC \cite{European_Environment_Agency_2011} is a valid open access alternative. The post-failure changes in vegetation cover were used within the model for estimating the effect of mass movement on soil erosion. 
Because of the modular modelling architecture (Fig. 2), the module that calculates the pre-failure C factor can be used as a link among our model and other approaches for measuring different land disturbance effects, in order to measure their effects on soil erosion.

The post-failure vegetation cover results were only partially altered by the slow mass movements that characterize this catchment (see Fig. 1). As locally the slide surface may also remain unchanged, we introduced into the model a value representing the post-failure percentage of bare soil. By analysing the landslide dataset, the available pictures, satellite images and accounting for all the information collected during a field survey carried out within the study area, the percentage of the post-failure bare soil cover was estimated to be not less than 20\% of the landslide area. For each of the pixels of the modelled landslides in each of the 1,000 Monte Carlo iterations, the \checkisnote{scalar\_positive} \checkisnote{proportion} of bare soil was therefore randomly determined in the range 0.2 - 1.

\section{Results and discussion}

Table 1 shows the results of the Monte Carlo simulations. We replaced the mean values obtained by applying equation 4, with the median, because it is more stable in that it is only marginally affected by extreme values. 
By analysing the median on 1,000 simulations of the cumulated pre-failure and post-failure soil erosion, an increase of 20\% of the total soil loss was estimated. The post-failure soil erosion rate in areas where landslides occurred is, on average, around 3.5 times the pre-failure value.

A bootstrap analysis based on 10,000 runs was performed in order to assess uncertainty.  The analysis of the changes in the rate of soil erosion due to landslide occurrence shows post-failure increases in soil loss of approximately 1700 tons per year (bootstrap p $\le$ 0.05).  This corresponds to an increase of around 22\% of the total soil erosion. We also analysed the extension of the area affected by slope instability. The bootstrap analysis shows that in each simulation at least 76 hectares, corresponding to around 8.5\% of the catchment, are affected by landslide activity (bootstrap p $\le$ 0.05). 
By comparing this value with the area that presented slope instability in 2006 (around 55 hectares), the applied methodology seems to result in a slight overestimate. The graph in Figure 3 shows that Malamud’s distribution seems to underestimate the number of small landslides ($<$ 300 $m^2$). Nevertheless, the probability density distribution for the Rocchetta landslides from 2006 is in line with those reported by Malamud et al. \cite{Malamud_etal_2004} for precipitation triggered landslides that took place in Guatemala in 1998. The model is in its early developmental phase and fine tuning the fit of the Malamud distribution to small landslides should help to improve the model predictions. However, for better evaluating the limits or the robustness of the proposed inverse-gamma distribution or of a modified version, further data would be necessary. The bootstrap analysis, with 10,000 runs, performed on the measured data (Fig. 4) shows the uncertainty associated with a single year landslide dataset is too high to extrapolate different parameter values. A more detailed analysis based on datasets covering a longer time interval would help to improve the applied methodology. An additional source of error contributing to the predictions, which needs further investigation, arises from the selection of the model for estimating soil erosion and its running with limited data: thus there is considerable scope for errors in prediction to be strongly linked to this simplification. 

Because the capacity to estimate the changes in soil erosion from landslide activity is largely dependent on the quality of the available datasets, the applied methodology broadens the possibility of a quantitative assessment of these effects in data-poor regions. The obtained results, even considering a possible overestimation, confirm the important role of mass movements on soil erosion and the consequent necessity to better integrate these processes into soil erosion modelling.

\begin{table}
\caption{Table 1: Bootstrap analysis of the modelling results. The bootstrap analysis, based on 10000 runs, shows the bootstrap cumulated distribution of the pre-and post-failure soil erosion within the area affected by landslide activity.}
\center\vspace{3mm}
\begin{tabular}{p{2.4cm}p{2.5cm}p{2.5cm}p{3.0cm}}
\hline \\[-3mm]
Quantile & 
Pre-failure \phantom{\hspace{8mm}} soil loss (t)&
Post-failure \phantom{\hspace{6mm}} soil loss (t)&
Estimated landslide activity area (ha)\\[1mm]
\hline \\[-2mm]
 5\% & 744.7 & 2530.3 & 76.6 (8.4\%)  \\
25\% & 799.2 & 2762.3 & 84.4 (9.2\%)  \\
50\% & 828.7 & 2773.3 & 85.5 (9.4\%)  \\
75\% & 843.4 & 2896 & 87.1 (9.6\%)   \\
95\% & 854.6 & 3005 & 88.9 (9.8\%)   \\[1mm]
\hline \\[-2mm]
 
\end{tabular}
\label{tab:bootstrapQuantiles}
\end{table}

\section{Conclusions}

A new method for empirically estimating the importance and extent of landslides on soil erosion losses in data-poor regions has been developed. This has been achieved by sampling the frequency-size landslide distribution proposed by Malamud et al. \cite{Malamud_etal_2004}, and stochastically distributing the landslide location across the catchment. Given the increasing threat of soil erosion all over the world and the implications this has on future food security and soil and water quality, an in-depth understanding of the rate and extent of soil erosion processes is crucial. 

Each year, on average, between 8.5 and 10\% of the catchment shows evidence of landslide activity that is responsible for a mean increase in the total soil erosion rate between 22 and 26\% above the pre-failure estimate. These results confirm the potential importance of integrating the landslide contribution into soil erosion modelling.
While this approach clearly has limitations the proposed approach can be seen as a first attempt to assess the landslide-erosion interaction in areas with limited data.

The proposed modelling approach is also suitable to be applied in applications having a wider spatial extent and to be potentially implemented in a transdisciplinary context. For example, the relevant effect of wildfires on soil erosion and landslide susceptibility \cite{Di_Leo_etal_ISESS2013,de_Rigo_etal_wildfireISESS2013} could be modelled with a higher reliability integrating the proposed approach. As stated in de Rigo et al. \cite{de_Rigo_etal_wildfireISESS2013}, wildfires can considerably increase soil erosion by water and landslide susceptibility. The changes in landslide susceptibility may in turn affect soil erosion.
In general, considering the modelling architecture (Fig. 2), if the module that calculates the pre-failure C factor value would provide the layer altered by a different disturbance (e.g. wildfires or outbreak of pests), the presented modelling architecture could be applied for estimating the indirect effect of these disturbances on soil erosion, provided a new landslide susceptibility map, that considers the altered vegetation cover, is produced .

Although the preliminary results are promising, further research is required before this method can be applied by the scientific community and relevant authorities with any level of confidence. Consideration of, and integrating within the model, post-failure changes in topography and soil characteristics (e.g. soil armouring \cite{Acharya_Cochrane_2008}) is fundamental for increasing the predictive capacity of the model. Also a better estimation of the bare soil exposed within a landslide is fundamental for improving our model. It would also be worthwhile to improve the fit of the Malamud distribution to the data that, at the present, it is not possible due to the limited availability of measured data. For obtaining more reliable results, and more robust estimates of the effects of landslides on soil and vegetation cover, it will be also necessary to focus attention on producing a less uncertain zonation of the spatial probability of the landslide susceptibility in areas characterized by low data availability \cite{Bosco_etal_2013}.

\subsection*{Acknowledgements}

We would like to thank Dr. Tom Dijkstra for his valuable comments on the manuscript. We also would like to thank Dr. Janusz Wasowski and Dr. Caterina Lamanna for providing the landslide data and Dr. Wasowski for his fundamental support during fieldwork. This paper is published with the support of the Maieutike Research Initiative.

\section*{Authors Bio}

\underline{Claudio Bosco} graduated in 2002 from the University of Milan with a degree in natural sciences. His more recent research activities are focused on natural hazards and their link with climate change, combining research into quantitative, robust modelling approaches with expert-driven understanding of environmental processes. His research interests also cover quantitative geomorphology, spatial analysis (GIS based) and wildfire effects on soil degradation processes.

\underline{Graham Sander} is professor of hydrology in the School of Civil and Building Engineering at Loughborough University in the UK.  His research interests cover sediment transport and soil erosion modelling, shallow overland flow, unsaturated subsurface water and contaminant transport.

\end{document}